\documentstyle[prl,aps,twocolumn]{revtex}

\begin{document}
\twocolumn[\hsize\textwidth\columnwidth\hsize\csname@twocolumnfalse\endcsname
\title{QUANTIZED SCALING OF GROWING SURFACES}
\author{Michael L\"assig}
\address{
Max-Planck-Institut f\"ur Kolloid- und Grenzfl\"achenforschung,
Kantstr.~55, 14513 Teltow, Germany}

\date{\today }
\maketitle

\begin{abstract}
The Kardar-Parisi-Zhang universality class of stochastic surface growth
is studied by exact field-theoretic methods. From previous numerical
results, a few qualitative assumptions are inferred. In particular,
height correlations should satisfy an operator product expansion and,
unlike the correlations in a turbulent fluid, exhibit no multiscaling.
These properties impose a quantization condition on the roughness
exponent $\chi$ and the dynamic exponent $z$.  Hence the exact values
$\chi = 2/5, z =  8/5$ for two-dimensional and $\chi = 2/7, z = 12/7$
for three-dimensional surfaces are derived.
\vspace{10 pt}

\vspace{24pt}
\end{abstract}


\vfill
] 
\narrowtext 

Strongly driven dynamic systems offer some of the most
intriguing realizations of statistical scale invariance. Hydrodynamic
turbulence~\cite{McComb} or the growth of rough
surfaces~\cite{KPZreviews} are two classic examples,
which turn out to be deeply connected from a theoretical point of
view.  In such systems, a stochastic force $\eta({\bf r},t)$
generates long-ranged correlations of a fluctuating dynamic field --
the local velocity ${\bf v}({\bf r},t)$ of a fluid or the height
$h({\bf r},t)$ of a surface. As typical differences
$h({\bf r}_ 1,t) - h({\bf r}_ 2,t)$ or ${\bf v}({\bf r}_ 1,t) -
{\bf v}({\bf r}_ 2,t)$ {\em increase} with the spatial separation
$|{\bf r}_ 1 - {\bf r}_ 2|$, the scaling properties of these fields are
generically more complex than those at a standard critical point.
Indeed, the theoretical understanding of these universality classes
far from equilibrium is still fragmentary.

The subject of this Letter is the simplest nonlinear model of stochastic
surface growth, the famous Kardar-Parisi-Zhang equation
\begin{equation}
\partial_t h  =   \nu \, {\bf \nabla}^2 h
                         + \frac{\lambda}{2} \, ( {\bf \nabla} h)^2 \,
                         + \eta
\label{KPZ}
\end{equation}
for a $d$-dimensional surface~\cite{KPZ}. The driving term
$\eta({\bf r},t)$, which describes the random adsorption of molecules onto
the surface, is taken to be Gauss distributed with correlations
only over {\em microscopic} distances,
\begin{equation}
\overline{\eta({\bf r},t) \eta({\bf r}',t')} = \sigma^2 \delta (t - t')
                                     \delta ({\bf r} - {\bf r}') \;.
\label{eta1}
\end{equation}
The relation of this model to the theory of turbulence is manifest:
Eq.~(\ref{KPZ}) is formally equivalent to Burgers' equation
\begin{equation}
\partial_t {\bf v} - ({\bf v} \cdot {\bf \nabla}) {\bf v}
          = \nu \, {\bf \nabla}^2 {\bf v}  + {\bf \nabla} \eta
\label{Burgers}
\end{equation}
for the driven dynamics of a vortex-free velocity field ${\bf v}({\bf r},t) =
{\bf \nabla} h({\bf r},t)$ (with $\lambda = 1$)~\cite{FNS}.
In a fluid, however, the driving force
is correlated over {\em macroscopic} spatial distances.
This leads to important differences in the scaling behavior~\cite{BMP.dp},
which are discussed below.

A surface growing from a flat initial state $h({\bf r}, 0) = 0$
develops height correlations with an increasing
correlation length $\xi_t \sim t^{1/z}$, which defines the dynamic
exponent $z$~\cite{L}.
A self-similar growth pattern,
characterized e.g. by the height difference moments
\begin{equation}
\langle [h({\bf r}_ 1) - h({\bf r}_ 2)]^k \rangle
      \sim |{\bf r}_ {12}|^{- k \chi }
\label{hk}
\end{equation}
(with ${\bf r}_ {12} \equiv {\bf r}_ 1 - {\bf r}_ 2$),
emerges on mesoscopic scales
$\tilde a \ll |{\bf r}_ {12}| \ll \xi_t $. For
$|{\bf r}_ {12}| \,\raisebox{-.6ex}{$\stackrel{\displaystyle <}{\sim}\,$}
 \tilde a$,
the dissipation term $\nu \, \nabla^2 h$ in Eq.~(\ref{KPZ}) breaks
the asymptotic scale invariance. In the scaling
regime (\ref{hk}), the height difference moments become {\em stationary},
i.e., independent of the correlation length $\xi_t$.
They are characterized by a single critical index $\chi \geq 0$, the {\em
roughness
exponent} of the surface. The scaling relation $\chi + z = 2$ follows from the
Galilei invariance of Eq.~(\ref{KPZ})~\cite{FisherHuse.dp}.
For $d = 1$, one can show Eq.~(\ref{hk})
to be valid with the roughness exponent $\chi = 1/2$, equal
to that of the linear theory
($\lambda = 0$)~\cite{KPZreviews}.
In higher dimensions, however, little is known analytically.
For $d > 2$, the rough state of the surface exists only if the
rescaled driving amplitude
$\lambda_0^2 \equiv \sigma^2 \lambda^2 / \nu^3$ exceeds a finite
threshold value $\lambda_c^2$~\cite{dim}. Less rigorous
theoretical arguments predict an upper critical dimension $d_> \leq 4$
beyond which Kardar-Parisi-Zhang surfaces are only
logarithmically rough ($\chi = 0$) even in the strong-coupling
regime $\lambda_0^2 > \lambda_c^2$~\cite{dupper}.
The numerical results presently available are
consistent with Eq.~(\ref{hk}). Extensive simulations yield
$\chi \approx 0.39$ for $d = 2$, $\chi \approx 0.31$ for $d = 3$, and smaller
positive values in higher dimensions, which are less reliable~\cite{num}.

It has remained a challenge for theorists to calculate the rough asymptotic
state of Kardar-Parisi-Zhang surfaces for $d > 1$ exactly or in a controlled
approximation. In particular, standard perturbative renormalization
about the linear theory fails to produce a fixed point
belonging to this regime~\cite{ren} -- a notorious difficulty familiar from the
theory of turbulence. In this Letter, a quite different approach is
taken. Guided by numerical and experimental results, I make a few
{\em qualitative} assumptions, namely the
existence of an operator product expansion~(\ref{ope}) and of a
stationary state (\ref{vk}) that is directed (i.e., has no up-down
symmetry). These assumptions turn out to constrain
severely the possible solutions of Eq.~(\ref{KPZ}). In particular, they
naturally lead to a quantization condition for the roughness exponent:
\begin{equation}
\chi = \frac{2}{k_0 + 2} \;,
\label{quant}
\end{equation}
where $k_0$ is an odd integer for $d \geq 2$.
Comparing with the above numerical estimates~\cite{errors}
and using the relation $\chi + z = 2$ then gives the main result
of this Letter: the exact values
$\chi = 2/5, z =  8/5$ for $d = 2$ and
$\chi = 2/7, z = 12/7$ for $d = 3$.

The fundamental observables describing the equal-time surface configurations
are the (connected) correlations
\begin{equation}
\langle h({\bf r}_ 1) \dots h({\bf r}_ n) \rangle_t
 = \int \! {\cal D}h  \, h({\bf r}_ 1) \dots h({\bf r}_ n) \,
  P_t \; - \dots
\label{hn}
\end{equation}
(the dots denoting the disconnected parts).
The height probability distribution $P_t (\{ h \})$ obeys the functional
Fokker-Planck equation
\begin{equation}
\partial_t P_t = \left. \int  d {\bf r} \left [
               \sigma^2 \frac{\delta^2}{\delta h({\bf r})^2}
               - \frac{\delta}{\delta h({\bf r})} J({\bf r})
               \right ] P_t \right. \;,
\label{Pt}
\end{equation}
where
$J({\bf r}) \equiv \nu {\bf \nabla}^2 h({\bf r}) +
   (\lambda/2) ({\bf \nabla} h)^2 ({\bf r})$
is the deterministic part of the current.

In the scaling regime ($\tilde a \ll |{\bf r}_ {ij}| \ll \xi_t$ for $i,j = 1,
\dots, n$),
the correlation functions (\ref{hn}) will generically
become singular as some of the points approach each other.
For $d < d_>$, these singularities are assumed to follow from
an {\em operator product expansion}
\begin{eqnarray}
 h({\bf r}_ 1) \dots h({\bf r}_ k) & = &
            \sum_{\cal O} |{\bf r}_ {12}|^{-k x_h + x_{\cal O}} \,
\nonumber \\
 & & \times \, C_k^{\cal O} \!\!\left ( \frac{{\bf r}_ {13}}{|{\bf r}_ {12}|},
\dots,
     \frac{{\bf r}_ {1k}}{|{\bf r}_ {12}|} \right ) \,
     {\cal O} ({\bf r}_ 1) \;.
\label{ope}
\end{eqnarray}
This identity is nothing but a consistency relation for the height
correlations. Inserted in (\ref{hn}), it
 expresses any $n$-point function as a sum
of $(n - k + 1)$-point functions in the limit
$|{\bf r}_ {ij}| \ll |{\bf r}_ {il}| \ll \xi_t$
($i,j = 1, \dots, k$ and
$l =  k+1, \dots, n$). The notion of an operator product expansion is familiar
in field theory~\cite{Cardy} and has recently been applied successfully
to nonequilibrium  systems~\cite{Polyakov,dep}.
The sum on the r.h.s. runs over all local scaling fields ${\cal O}({\bf r})$.
Each term
contains a dimensionless scaling function $C_k^{\cal O}$ (a simple number
for $k = 2$) and a power of $|{\bf r}_ {12}|$ given by the scaling dimensions
$x_{\cal O}$ and $x_h = - \chi$ (such that the overall dimension equals
that of the l.h.s.). The field ${\cal O}_k$ with the smallest dimension,
$x_k$, determines in
particular the asymptotic behavior of the $k$-point functions as
$\xi_t \to \infty$,
\begin{equation}
\langle h({\bf r}_ 1) \dots h({\bf r}_ k) \rangle_t
        \sim \langle {\cal O}_k \rangle_t
        \sim \xi_t^{-x_k} \;.
\label{hk2}
\end{equation}
The amplitudes
$\langle {\cal O}_k \rangle_t = \langle {\cal O}_k ({\bf r}) \rangle_t$
diverge with $\xi_t$, i.e., $x_k < 0$~\cite{normalordering}.
They measure the {\em global} roughness, which increases as the
surface develops higher mountains and deeper valleys.
{\em Local} surface properties should,
however, behave quite differently. For example, the gradient correlation
functions are assumed to have a finite limit
\begin{equation}
\lim_{\xi_t \to \infty}
 \langle {\bf \nabla} h({\bf r}_ 1) \dots
         {\bf \nabla} h({\bf r}_ n) \rangle_t
 \equiv \langle {\bf \nabla} h({\bf r}_ 1) \dots
                {\bf \nabla} h({\bf r}_ n) \rangle \;.
\label{vk}
\end{equation}
By writing
$h({\bf r}_ i) - h({\bf r}_ i') =
 \int_{{\bf r}_ i}^{{\bf r}_ i'}  {\bf ds} \cdot {\bf \nabla} h ({\bf s})$, the
same
property follows for the height difference correlation functions
$\langle \prod_{i = 1}^n [h({\bf r}_ i) - h({\bf r}_ i')] \rangle_t$,
in particular for the moments (\ref{hk}). This implies a feature familiar from
simulations: one cannot recognize the value of $\xi_t$ from snapshots
of the surface in a region much smaller than $\xi_t$.

The operator product expansion (\ref{ope}) induces an expansion for the
gradient field ${\bf v} \equiv {\bf \nabla} h$ of the form
\begin{eqnarray}
{\bf v}({\bf r}_ 1) \dots {\bf v}({\bf r}_ k) & = &
      \sum_{\cal O} |{\bf r}_ {12}|^{- k x_{\bf v} + x_{\cal O}} \,
\nonumber \\
& &   \times \, \tilde C^{\cal O}_k
      \!\!\left ( \frac{{\bf r}_ {13}}{|{\bf r}_ {12}|}, \dots,
      \frac{{\bf r}_ {1k}}{|{\bf r}_ {12}|} \right )
      {\cal O}({\bf r}_ 1)
\label{vope}
\end{eqnarray}
with new scaling functions $\tilde C_k^{\cal O}$ and the dimension
$x_{{\bf v}} = - \chi + 1$. (Both sides of
(\ref{vope}) are tensors of rank $k$ whose indices are suppressed.)
Hence, the stationarity condition (\ref{vk}) allows only
two kinds of terms in (\ref{ope}):
\newline
(a) {\em singular} terms involving fields ${\cal O}({\bf r})$ with a
{\em non-negative} scaling dimension such as ${\bf 1}$ (the identity field),
$({\bf \nabla} h)^2({\bf r})$, etc. Any such field that appears in (\ref{ope})
will
also appear in (\ref{vope}) and generate a term
$\langle {\bf v}({\bf r}_ 1) \dots {\bf v}({\bf r}_ k) \rangle_t
  \sim \langle {\cal O} \rangle_t
  \sim \xi_t^{-x_{\cal O}}$
that violates (\ref{vk}) if $x_{\cal O} < 0$~\cite{fourpoint}.
\newline
(b) {\em regular} terms, where the coefficient
$|{\bf r}_ {12}|^{-k x_h + x_{\cal O}} \, C_k^{\cal O}$
is a tensor of rank $N$ in the differences ${\bf r}_ {1i}$ ($i = 2,
\dots, k)$. Such terms do not violate (\ref{vk}) since
they have a vanishing coefficient $\tilde C_k^{\cal O}$ in (\ref{vope})
for $N < k$. The leading ($N = 0$) term
involves the (normal-ordered) field
${\cal O}_k ({\bf r}) = h^k  ({\bf r})$ and governs
the asymptotic singularity (\ref{hk2}); the higher
terms correspond to fields with $k$ factors $h({\bf r})$ and $N$ powers of
${\bf \nabla}$. These composite fields have dimensions
\begin{equation}
x_{k,N} = - k \chi + N \;.
\label{xkN}
\end{equation}

It is useful to introduce the (normal-ordered) vertex fields
$Z_q ({\bf r}) \equiv \exp [q h({\bf r})]$, which are the generating functions
of the fields $h^k ({\bf r})$. Eq.~(\ref{ope}) then implies
the operator product expansion
\begin{eqnarray}
Z_{q_1} ({\bf r}_ 1) Z_{q_2} ({\bf r}_ 2) & = &
     {\rm exp} \! \left ( \sum_{k,l} C_{k,l}^{\bf 1} w_1^k w_2^l \right )
     Z_{q_1 + q_2} ({\bf r}_ 1)
\nonumber \\
& & + \; O ( C_{k,l}^{{\cal O} \neq 1}) \;,
\label{Zope}
\end{eqnarray}
where  $C_{k,l}^{\cal O} \equiv
      C_{k+l}^{\cal O} (0,\dots, 0,
      {\bf r}_ {12}/|{\bf r}_ {12}|, \dots, {\bf r}_ {12}/|{\bf r}_ {12}|)$
with the first $k$ arguments equal to 0 and
$w_i \equiv q_i |{\bf r}_ {12}|^\chi$~\cite{long}. Subleading singular terms
(with positive-dimensional fields ${\cal O}$) and regular terms
(with fields containing height gradients) are omitted.
The vertex $n$-point functions
$\langle Z_{q_1} ({\bf r}_ 1) \dots Z_{q_n} ({\bf r}_ n) \rangle_t$ behave
asymptotically as $\exp (\xi_t^\chi \sum_{i = 1}^n q_i)$. If $\sum_i q_i = 0$,
they  have a finite limit
$\langle Z_{q_1} ({\bf r}_ 1) \dots
         Z_{- q_1 \dots - q_{n-1}} ({\bf r}_ n) \rangle$. Since
these are precisely the vertex correlators that generate the height
difference correlation functions and since (\ref{Zope}) is analytic in
the $q_i$, this leads back to the stationarity condition (\ref{vk}).

The operator product expansion (\ref{Zope}) with the linear dimensions
(\ref{xkN}) is at the heart of the field theory for Kardar-Parisi-Zhang
systems. It is instructive to compare this theory with models of
turbulence. Burgers' equation (\ref{Burgers}) with force correlations
\begin{equation}
\overline{\eta({\bf r},t) \eta({\bf r}',t')}
     = \sigma^2 R^2 \delta (t - t') \Delta(|{\bf r} - {\bf r}'|/R)
\end{equation}
over large distances $R$ develops {\em multiscaling}: for example,
the longitudinal velocity difference moments
\begin{equation}
\langle [v_{\|} ({\bf r}_ 1) - v_{\|} ({\bf r}_ 2)]^k \rangle \sim
   |{\bf r}_ {12}|^{-k x_{{\bf v}} + \tilde x_k} \, R^{-\tilde x_k}
\label{multisc}
\end{equation}
have a $k$-dependent singular dependence on $|{\bf r}_ {12}|$ and $R$
in the inertial scaling regime
$\tilde a \ll |{\bf r}_ {12}| \ll R$~\cite{ChekhlovYakhot,BMP.dp}.
Similar multiscaling is present in Navier-Stokes turbulence.
Kolmogorov's famous argument predicts the exact scaling dimension of
the velocity field, $x_{{\bf v}} = -1/3$, from dimensional analysis~\cite{K41}.
This
determines the scaling of the third moment in (\ref{multisc})
since $\tilde x_3 = 0$. The higher exponents $\tilde x_4, \tilde x_5, \dots <
0$ cannot
be obtained from dimensional analysis. Assuming the existence of
an operator product expansion (\ref{vope}), the term (\ref{multisc}) is
generated by the lowest-dimensional field $\tilde {\cal O}_k$
with a singular coefficient~\cite{Eyink}. Multiscaling thus implies
the existence of a (presumably infinite) number of composite fields with
anomalous negative dimensions. For the velocity vertex fields
${\rm exp} [q v(r)]$ of Burgers turbulence in one dimension,
Polyakov has conjectured an operator
product expansion similar to (\ref{Zope})~\cite{Polyakov}.
It is no longer analytic in $q_1, q_2$  and thus consistent with multiscaling.
The distinguishing feature of
Kardar-Parisi-Zhang surfaces is the absence of multiscaling~\cite{Krug}.
Notice that
the resulting properties (\ref{xkN}) and (\ref{Zope}) have been derived
solely from the assumptions (\ref{ope}) and (\ref{vk})
without using Eq.~(\ref{KPZ}) explicitly.

To establish the consistency of the
operator product expansion with the underlying dynamic equation,
one has to construct correlation functions that remain finite
in the continuum limit $\tilde a \to 0$. With the probability
distribution (\ref{Pt}), the height correlations (\ref{hn}) develop
singularities dictated by their normalization in the linear regime
($|{\bf r}_ {ij}| \ll \tilde a$). The existence of a well-defined asymptotic
scaling
regime for $|{\bf r}_ {ij}| \gg a$ implies that these singularities can be
absorbed by a change of variables
\begin{equation}
h({\bf r}) \to {\cal Z}_h (\tilde a / r_0) \, h ({\bf r})
\;, \hspace{0.5cm}
t \to {\cal Z}_t (\tilde a / r_0) \, t
\label{Z}
\end{equation}
such that the ``renormalized'' correlations (\ref{hn})
satisfy normalization conditions independently of $\tilde a$
at some mesoscopic scale $r_0$~\cite{Lassig.KPZ,long}.
The Z-factors have the asymptotic behavior
${\cal Z}_h \sim (\tilde a / r_0)^{\chi - \chi_0}$ and
${\cal Z}_t \sim (\tilde a / r_0)^{z - z_0}$ as
$\tilde a/r_0 \to 0$,
where $\chi_0 = (2 - d)/2$ and $z_0 = 2$ are the exponents in the
linear regime.
Of course, I do not assume perturbative renormalizability (i.e., that
the Z-factors are analytic functions of $\lambda_0^2$). Since the
scaling dimensions (\ref{xkN}) are linear in $k$, the renormalization
(\ref{Z}) also removes the singularities from correlations of the
fields $h^k ({\bf r})$ and $Z_q({\bf r})$, ensuring a finite limit
of the coefficients $C$ in (\ref{ope}), (\ref{vope}), (\ref{Zope})
and of the amplitudes
$\langle {\cal O}_k \rangle_t$ in (\ref{hk2}). The substitution
(\ref{Z}) also leads to new coefficients in (\ref{KPZ}) and~(\ref{Pt}):
\begin{equation}
\begin{array}{rcl}
\nu (\tilde a / r_0) & \sim &
       {\cal Z}_t^{-1}
       \simeq \nu^* \times (\tilde a / r_0)^\chi \;,
\\
\sigma^2 (\tilde a / r_0) & \sim &
       {\cal Z}_h^2 {\cal Z}_t^{-1}
       \simeq \sigma^{* 2} \times (\tilde a / r_0)^{d - 2 + 3 \chi} \;,
\\
r_0^{\chi_0} \lambda (\tilde a, r_0) & \sim &
       {\cal Z}_t^{-1} {\cal Z}_h^{-1}
       \simeq g^* \;.
\end{array}
\label{coeff}
\end{equation}
Galilei invariance is expressed by the asymptotic scale invariance of
the dimensionless coupling $r_0^{\chi_0} \lambda$, while the other
coefficients become irrelevant as $\tilde a / r_0 \to 0$.
However, as explained in ref.~\cite{Polyakov} for Burgers turbulence, the
equation of motion for the renormalized correlation functions
is quite subtle due to anomalies dictated by the operator product expansion.
To exhibit the anomalies for the height correlations, I
introduce the smeared vertex fields
$
Z_{q}^a ({\bf r}) \equiv
       \exp [q  \int d {\bf r}' \, \delta_a ({\bf r} - {\bf r}') h({\bf r}')  ]
$
(where $\delta_a ({\bf r})$ is a normalized function with support in the
sphere $|{\bf r}| < a$) and the abbreviations
$Z_i^a \equiv Z_{q_i}^a ({\bf r}_ i),
 Z_i   \equiv Z_{q_i}   ({\bf r}_ i)$.
Using (\ref{hn}), (\ref{Pt}), and (\ref{coeff}), it
is straightforward to derive
\begin{equation}
\partial_t \, \langle Z_1^a \dots Z_n^a \rangle_t =
   \sum_{i = 1}^n q_i
   \langle Z_1^a \dots {\cal J} \! Z_i^a \dots Z_n^a \rangle_t \;,
\label{Znta}
\end{equation}
where
$ {\cal J} \! Z_i^a \equiv [q_i \sigma^2 \delta_a (0) + J({\bf r}_ i)] Z_i^a$.
The singularity structure of the current is determined by
(\ref{ope}) and (\ref{coeff}):
\begin{equation}
{\cal J}\! Z_i^a = g^* \hat Z_i + a^{2 \chi - 2}
                   \left ( \sum_{k = 1}^{\infty} c_k a^{k \chi} q_i^k
                   \right ) Z_i + O(a^\chi)
\label{JZ}
\end{equation}
for $a, \tilde a \to 0$ with $a / \tilde a$ kept constant.
The field
$\hat Z_q ({\bf r}) \equiv (\nabla h)^2 Z_{q} ({\bf r})$ denotes the finite
part of the operator product
$(\nabla h)^2 ({\bf r}) Z_q^a ({\bf r})$ for $a \to 0$, and
$\hat Z_i \equiv \hat Z_{q_i} ({\bf r}_ i)$.
The finite dissipation term $(\nabla^2 h) Z_{q_i} ({\bf r}_ i)$
becomes irrelevant in this limit since $\nu \sim a^\chi$.
The singular part of (\ref{JZ}) is a power series
in $q_i$ with asymptotically constant coefficients
$c_1 = \sigma^{* 2} a^d \delta_a (0)
       + \nu^* c_{1,1} + g^* c_{2,1}$ and
$c_k = \nu^* c_{1,k} + g^* c_{2,k}$ for $k = 2,3,\dots$.
The terms of order $a^{(2 + k) \chi - 2}$ originate from operator products
$\nabla^2 h ({\bf r}_ i) h ({\bf r}_ 1') \dots h ({\bf r}_ k') \sim {\bf 1}$
and
$(\nabla h)^2 ({\bf r}_ i) h ({\bf r}_ 1') \dots h ({\bf r}_ k') \sim {\bf 1}$;
their
respective coefficients $c_{1,k}$ and $c_{2,k}$ are integrals
over the scaling functions in (\ref{ope}) and the regularizing functions
$\delta_a ({\bf r}_ i - {\bf r}_ j')$. Of course, divergent terms have to
cancel
so that Eq.~(\ref{Znta}) has a finite continuum limit
\begin{equation}
\partial_t \, \langle Z_1 \dots Z_n \rangle_t =
   \sum_{i = 1}^n q_i
   \langle Z_1 \dots {\cal J} \! Z_i \dots Z_n \rangle_t
\label{Znt}
\end{equation}
with ${\cal J} \! Z_i = \lim_{a \to 0} {\cal J} \! Z_i^a$.
For generic values of $\chi$,
this implies $ {\cal J} \! Z_i = g^* \hat Z_i$. However, if $\chi$ satisfies
the condition (\ref{quant}) for some integer $k_0$, the dissipation
current contributes an anomaly:
\begin{equation}
{\cal J} \! Z_i = g^* \hat Z_i + \nu^* c_{1,k_0} q_i^{k_0} Z_i \;.
\label{anomaly}
\end{equation}

Eqs.~(\ref{Znt}) and (\ref{anomaly}) govern in particular the stationary
state of the surface. For $d = 1$, the stationary height distribution
is known,
$P \sim  {\rm exp}
         [- (\sigma^2 / \nu) \int d {\bf r} \, (\nabla h)^2]$.
It equals that of the linear theory, thus restoring the up-down symmetry
$ h({\bf r}) - \langle h \rangle_t \to
- h({\bf r}) + \langle h \rangle_t$ broken by
the nonlinear term in (\ref{KPZ}). The exponent $\chi = 1/2$ satisfies
(\ref{quant}) with $k_0 = 2$ but the up-down symmetry forces the anomaly
to vanish ($c_{1,2} = 0$). In higher dimensions, this symmetry is
expected to remain broken in the stationary regime. The surface has
rounded hilltops and steep valleys, just like the upper side of a cumulus
cloud~\cite{Pelletier}. Hence, the local slope is correlated with the relative
height, resulting in nonzero odd moments
$\langle (\nabla h)^2 ({\bf r}_ 1) [h({\bf r}_ 1) - h({\bf r}_ 2)]^k \rangle$.
However, this is consistent with Eqs.~(\ref{Znt}) and (\ref{anomaly})
only for odd values of $k_0$, where
\begin{eqnarray}
\lefteqn{\langle \hat Z_q ({\bf r}_ 1) Z_{-q} ({\bf r}_ 2) \rangle -
\langle \hat Z_{-q} ({\bf r}_ 1) Z_q ({\bf r}_ 2) \rangle}
\nonumber \\
& &   \hspace*{1.5cm}
    =  - (\nu^* / g^*) \, c_{1,k_0} q^{k_0}
       \langle Z_q ({\bf r}_ 1) Z_{-q} ({\bf r}_ 2) \rangle
\end{eqnarray}
and hence for odd values of $k \geq k_0$
\begin{eqnarray}
\lefteqn{
\langle (\nabla h)^2 ({\bf r}_ 1) [h({\bf r}_ 1) - h({\bf r}_ 2)]^k \rangle}
\nonumber \\
& &  \hspace*{1.5cm}
    = - (\nu^* / g^*) \, c_{1,k_0}
      \langle [h({\bf r}_ 1) - h({\bf r}_ 2)]^{k - k_0} \rangle \;.
\label{direct}
\end{eqnarray}
The directedness of the stationary growth pattern thus requires
a nonzero anomaly $c_{1,k_0}$ with an odd integer $k_0$.
The roughness exponent is then determined by Eq.~(\ref{quant}).
The values $k_0 = 3$ for $d = 2$ and $k_0 = 5$ for $d = 3$
give the exponents quoted above, in reasonable agreement with numerical
results~\cite{num,errors}.

In summary, the scaling of growing surfaces has been determined by
requiring consistency of the effective large-distance field theory
subject to a few phenomenological constraints. The Galilei symmetry of
the dynamic equation conspires with these constraints to
allow only discrete values of the roughness exponent in two and three
dimensions. The underlying solutions of Eq.~(\ref{KPZ})
are distinguished by a dynamical anomaly in the strong-coupling regime:
the dissipation term contributes a finite part to the effective equation
of motion (\ref{Znt}) despite being formally irrelevant. The anomaly
manifests itself in identities like (\ref{direct}) between stationary
correlation functions.
The quantization rule (\ref{quant})  is analogous to the
exact Kolmogorov scaling of the third velocity difference moment in
Navier-Stokes turbulence.  The deeper reason for this rigidity is yet
to be explained.

I am grateful to D. Wolf for useful discussions.


\begin{references}
\bibitem{errors}
The accuracy of the numerical data may be affected by corrections to
scaling not taken into account systematically. The corrections are
expected to vary with $d$ since the short-distance regime
($|{\bf r}_ {12}| \ll \tilde a$) is governed by the linear theory
($\lambda_0 = 0$) for $d \leq 2$ and by the critical point
$\lambda_0^2 = \lambda_c^2$ for $d > 2$.
\bibitem{normalordering}
The average local height has the  nonuniversal part
$\langle h \rangle_t = u_0 t$ independent of $\xi_t$.
It is eliminated by normal-ordering
Eq.~(\ref{KPZ}):
$ (\nabla h)^2 \to
  (\nabla h)^2 - \langle (\nabla h)^2 \rangle_{\infty} $.
The higher connected correlations (\ref{hn}) remain invariant.
\end{references}
\end{document}